\documentclass[prd,a4paper,twocolumn,preprintnumbers,nofootinbib]{revtex4-2}

\usepackage{amsmath,amssymb,slashed,mathtools,slashed}
\usepackage{graphicx,multirow}
\usepackage[hyperfootnotes=false,colorlinks,citecolor=blue]{hyperref}
\usepackage{mathtools,halloweenmath}

\newcommand{\beq}{\begin{equation}}
\newcommand{\eeq}{\end{equation}}
\newcommand{\bea}{\begin{eqnarray}}
\newcommand{\ena}{\end{eqnarray}}

\def \epsilon {\varepsilon} 

\usepackage{siunitx} 
\DeclareSIUnit{\barn}{b}

\usepackage{enumitem}

\newcommand{\mdzero}{m_{\delta^0}}
\allowdisplaybreaks

\begin{document}

\title{Minimal model for the $W$-boson mass, $(g-2)_\mu$, $h\to\mu^+\mu^-$ and quark-mixing-matrix unitarity}

\author{Andreas Crivellin}
\email{andreas.crivellin@cern.ch}
\affiliation{Physik-Institut, Universitat Zurich, Winterthurerstrasse 190, CH–8057 Zurich, Switzerland}
\affiliation{Paul Scherrer Institut, CH–5232 Villigen PSI, Switzerland}

\author{Matthew Kirk}
\email{mjkirk@icc.ub.edu}
\affiliation{Departament de Física Quàntica i Astrofísica (FQA),
Institut de Ciències del Cosmos (ICCUB), Universitat de Barcelona (UB), Spain}

\author{Anil Thapa}
\email{wtd8kz@virginia.edu}
\affiliation{Department of Physics, University of Virginia, Charlottesville, Virginia 22904-4714, USA}

\hypersetup{
pdftitle={Y=0 Scalar Triplet Beyond the W Mass: (g-2)mu, h -> mu mu and CKM Unitarity},   
pdfauthor={Anil Thapa, Andreas Crivellin, Matthew Kirk}
}

\preprint{PSI-PR-23-11, ZU-TH 20/23}

\begin{abstract}
The $SU(2)_L$ triplet scalar with hypercharge $Y=0$ predicts a positive definite shift in the $W$ mass, w.r.t.~the Standard Model prediction, if it acquires a vacuum expectation value. As this new field cannot couple directly to SM fermions (on its own), it has no significant impact on other low-energy precision observables and is weakly constrained by collider searches. In fact, the multi-lepton anomalies at the LHC even point towards new scalars that decay dominantly to $W$ bosons, as the neutral component of the triplet naturally does. In this article, we show that with a minimal extension of the scalar triplet model by a heavy vector-like lepton, being either I) an $SU(2)_L$ doublet with $Y=-1/2$ or II) an $SU(2)_L$ triplet with $Y=-1$, couplings of the triplet to Standard Model leptons are possible. This minimal extension can then provide, in addition to the desired positive shift in the $W$ mass, a chirally enhanced contribution to $(g-2)_\mu$. In addition version I) and II) can improve on $Z\to\mu^+\mu^-$ and alleviate the tension in first-row CKM unitarity (known as the Cabibbo angle anomaly), respectively. Finally, both options, in general, predict sizable changes of $h\to\mu^+\mu^-$, i.e.,~much larger than most other $(g-2)_\mu$ explanations where only $O(\%)$ effects are expected, making this channel a smoking gun signature of our model.
\end{abstract}

\maketitle

\section{Introduction}
\label{intro}
The Standard Model (SM) of particle physics is the theory describing the fundamental constituents and interactions of matter according to our current state of knowledge. However, it is clear that it cannot be the ultimate description of nature. For instance, it cannot account for the existence of Dark Matter established at cosmological scales, nor for the non-vanishing neutrino masses required by neutrino oscillations. Unfortunately, these observations can be addressed in many different ways and within a very wide range for the new physics scale. Therefore, in the absence of confirmed direct signals for new particles, more information on possible extensions of the SM is thus necessary to make progress towards a theory superseding the SM that can be tested at the Large Hadron Collider (LHC) or next-generation experiments. In this context, we can use deviations from the SM predictions in low-energy observables, known as anomalies, as a guide for identifying promising extensions of the SM, within which one can then calculate predictions for future verification (or falsification) of the model. Prominent candidates among these indirect hints for physics beyond the SM (see e.g.~Ref~\cite{Crivellin:2023gky} for a recent review) are the anomaly in the $W$ boson mass~\cite{CDF:2022hxs}, the anomalous magnetic moment of the muon ($(g-2)_\mu$)~\cite{Abi:2021gix} as well as the deficit in first-row CKM unitarity, known as the Cabibbo angle anomaly (CAA)~\cite{Crivellin:2022ctt}. While in the first observable, the CDF~II result is in some tension with LHC measurements~\cite{LHCb:2021bjt,ATLAS:2023fsi}, the significance of the deviation in $(g-2)_\mu$ depends on the SM prediction, where inconsistencies between the data-driven method~\cite{Aoyama:2020ynm} and lattice results~\cite{Borsanyi:2020mff} exist. However, it is still very interesting and instructive to see which models can give sizable effects in these observables. While several combined NP explanations of $(g-2)_\mu$ and the $W$ mass have been proposed in the literature~\cite{DAlise:2022ypp,Bagnaschi:2022qhb,Babu:2022pdn,Cheung:2022zsb,Kawamura:2022uft,Nagao:2022oin,Arcadi:2022dmt,Chowdhury:2022moc,Bhaskar:2022vgk,Baek:2022agi,Kim:2022hvh,Kim:2022hvh,Kim:2022xuo,Dcruz:2022dao,Chowdhury:2022dps,Chakrabarty:2022voz,Atkinson:2022qnl,Arora:2022uof,Hou:2022mwr,Yang:2022gvz,Domingo:2022pde,Cacciapaglia:2022evm,deGiorgi:2022xhr,Primulando:2022vip,Pfeifer:2022yrs,Chen:2023eof,Mishra:2023cjc,Ahmadvand:2023gse,Heeck:2023iqc,Han:2022juu}, a simple combined explanation of all three anomalies is still missing (to the best of our knowledge). 

In this article, we aim at constructing a minimal model that can naturally provide sizable effects in both $m_W$ and $(g-2)_\mu$ (and possibly explain the CAA) and investigate its phenomenological consequences. For this, our starting point is the $\Delta$SM~\cite{Ross:1975fq, Gunion:1989ci,Chankowski:2006hs,Blank:1997qa,Forshaw:2003kh, Chen:2006pb,Chivukula:2007koj,Bandyopadhyay:2020otm}, where an $SU(2)_L$ triplet scalar with hypercharge 0 ($\Delta$) is added to the SM particle content. Its vacuum expectation value (VEV) violates custodial symmetry at the tree-level via a positive contribution to the $W$ boson mass, as suggested by the measurement of the CDF~II measurement~\cite{Chabab:2018ert,FileviezPerez:2022lxp, Cheng:2022hbo, Chen:2022ocr,Rizzo:2022jti,Chao:2022blc,Wang:2022dte,Shimizu:2023rvi,Lazarides:2022spe,Senjanovic:2022zwy}. Since the neutral component of the triplet scalar can dominantly decay to pairs of $W$ bosons, while the decay to $Z$ pairs is suppressed by mixing with the SM Higgs, this model is well motivated by the LHC multi-lepton anomalies~\cite{vonBuddenbrock:2017gvy,Buddenbrock:2019tua,vonBuddenbrock:2020ter,Hernandez:2019geu,Fischer:2021sqw}, including the hint for an enhanced $W$ pair production at the electroweak (EW) scale~\cite{Coloretti:2023wng}.

\begin{figure}
    \centering
    \includegraphics[scale=0.35]{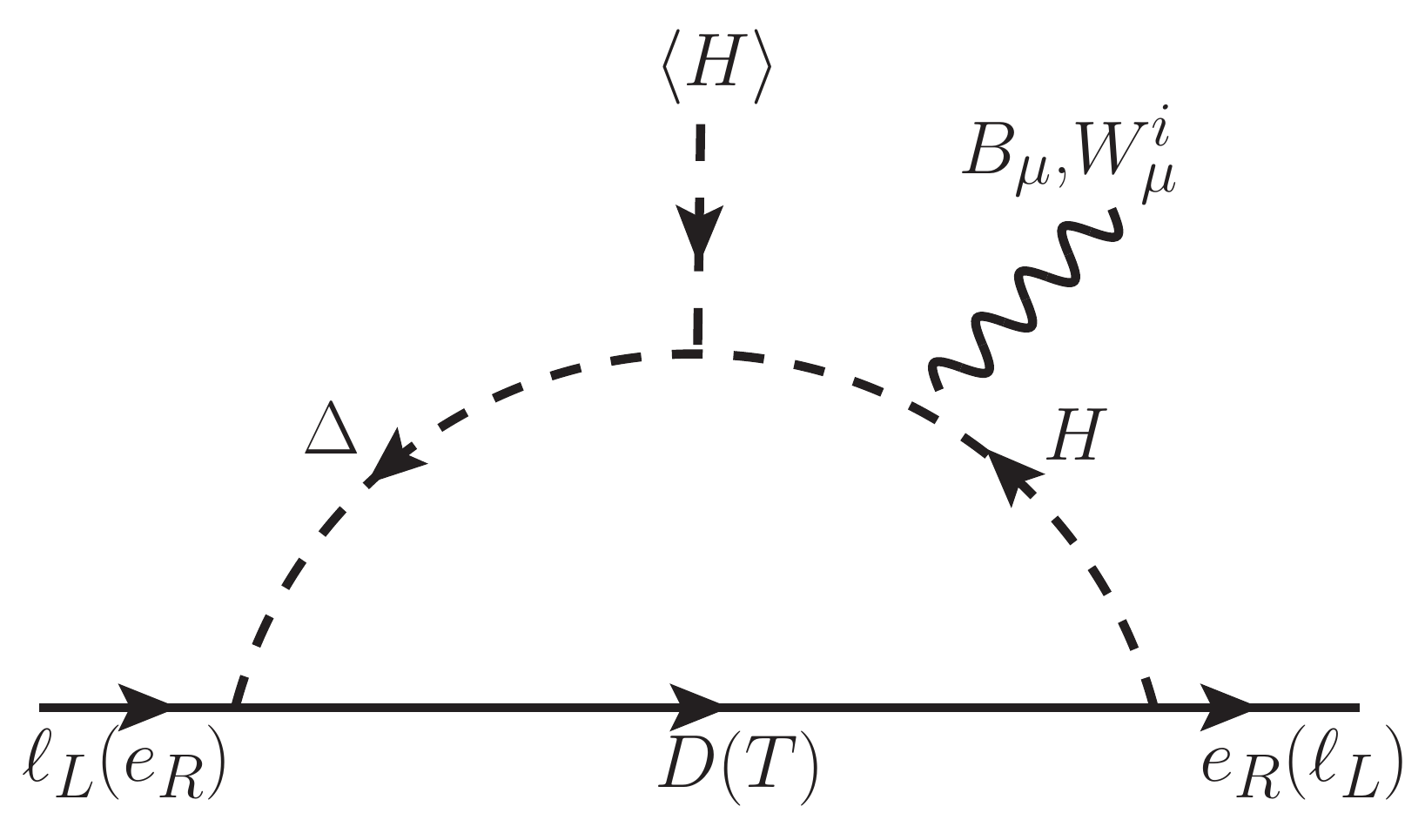} \\[5pt] 
    \includegraphics[scale=0.37]{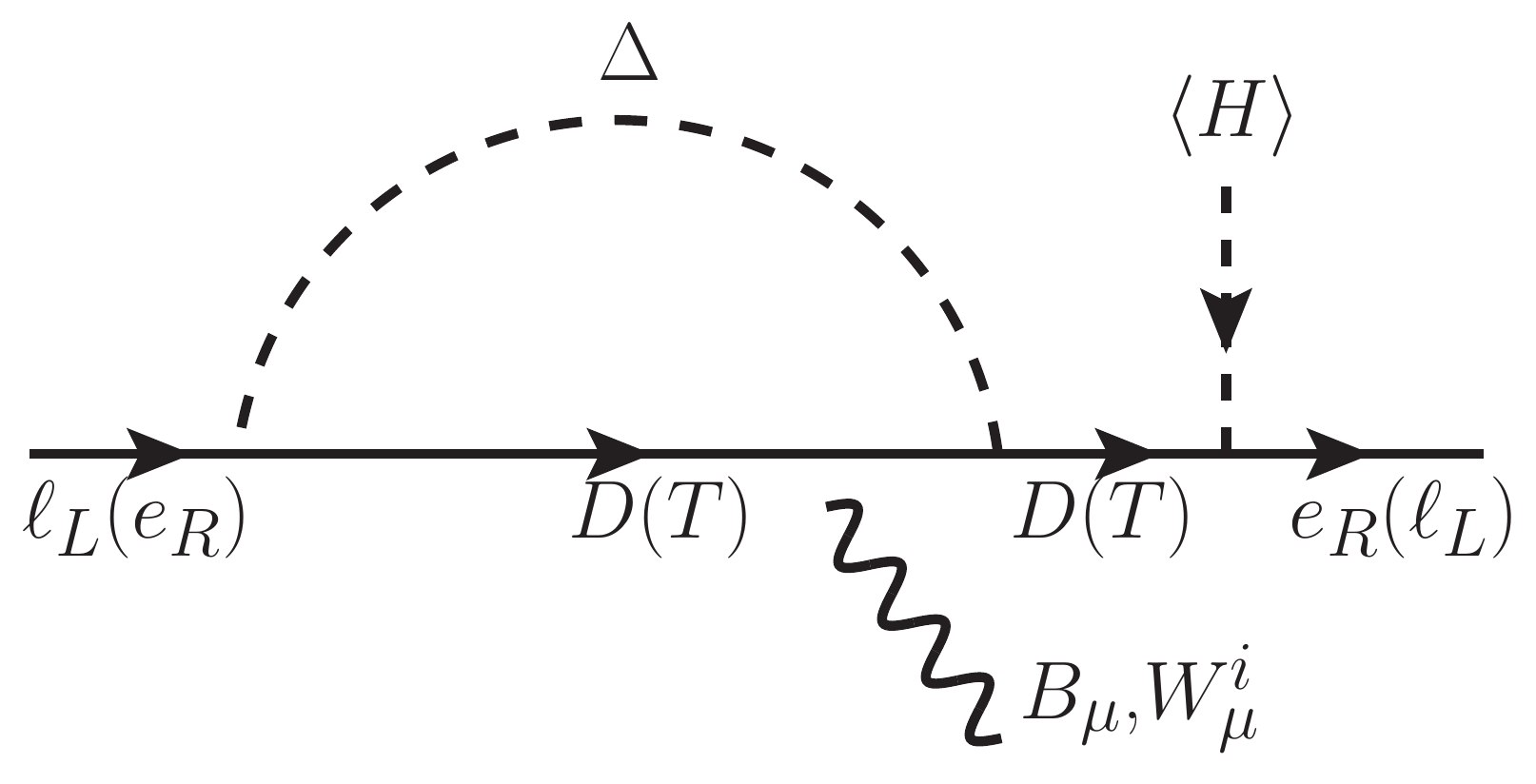} 
    \caption{Leading one-loop effect contribution to $(g-2)_\mu$ from the VLL doublet $D$ (triplet $T$). Note that while the upper diagram involves $\mu$, the lower diagram is proportional to the VEV of the SM doublet and therefore in general dominant. }
    \label{fig:1loopfer}
\end{figure}

Next, we aim at extending the $\Delta$SM to obtain a sizable effect in $g-2$ of the muon. In fact, there are only two minimal options that can, as we will show, give rise to chirally enhanced effects in $(g-2)_\mu$ (see e.g.~Refs.~\cite{Czarnecki:2001pv,Stockinger:2009fns,Giudice:2012ms,Freitas:2014pua,Kowalska:2017iqv,Crivellin:2018qmi,Capdevilla:2021rwo,Guedes:2022cfy} for generic models with chiral enhancement). We can supplement the $\Delta$SM by a vector-like lepton
\begin{equation}
    {\rm I)}\ D \sim (1,2,-1/2)\, \qquad {\rm or} \qquad
    {\rm II)}\ T \sim (1,3,-1) \, ,  \nonumber
\end{equation}
where the numbers in the bracket denote their representation under the SM gauge group $SU(3)_C\times SU(2)_L\times U(1)_Y$. The corresponding Feynman diagrams giving the dominant contribution to $g-2$ are shown in Fig.~\ref{fig:1loopfer}.\footnote{Note that contrary to Ref.~\cite{Crivellin:2021rbq}, or the MSSM with R-parity conservation (see Ref.~\cite{Stockinger:2006zn} for a review), we do not impose a (effective) $Z_2$ symmetry. This extension allows for a more minimal setup in which only two new fields are needed (instead of 3 in Ref.~\cite{Crivellin:2021rbq}).} 

Interestingly, both model versions lead to unavoidable tree-level effects in the dim-6 operator $(H^\dagger H)(\bar{\ell}_L H e_R)$~\cite{Grzadkowski:2010es} contributing to the muon mass and $h\to\mu^+\mu^-$ after electroweak (EW) symmetry breaking. In fact, while most other $(g-2)_\mu$ explanations only predict effects of the order of a few percent~\cite{Crivellin:2020tsz,Crivellin:2021rbq,Fajfer:2021cxa}, we will see that our model, in general, predicts much larger effects.

\section{Model}

Our starting point is to add a real scalar $SU(2)_L$ triplet with $Y=0$ ($\Delta$) to the SM particle content, called the $\Delta$SM. The most general renormalizable scalar potential involving SM Higgs $H$ and real triplet $\Delta$ reads
\begin{align}
    V =& -\mu_H^2 H^\dagger H + \mu_\Delta^2{\rm Tr}[\Delta^2] + \lambda_1(H^\dagger H)^2 + \lambda_2{\rm Tr}[\Delta^4] \nonumber \\
    & + \lambda_3(H^\dagger H)\, {\rm Tr}[\Delta^2]  + \mu H^\dagger \Delta H\,,
\end{align}
where the scalar fields are defined as
\begin{equation}
    H = 
    \begin{pmatrix}
    H^+ \\   H^0
    \end{pmatrix}, \hspace{5mm}
    \Delta = \frac{1}{2} \begin{pmatrix}
        \Delta^0 & \sqrt{2} \Delta^+ \\
        \sqrt{2} \Delta^- & - \Delta^0
    \end{pmatrix}\,,
\end{equation}
in terms of electric charge eigenstates. The scalar potential has a global $O(4)_H \times O(3)_\Delta$ symmetry in the limit $\mu \to 0$. Therefore, $\mu$ softly breaks this symmetry and is naturally small. We denote the VEVs as $\langle H^0 \rangle = {v}/{\sqrt{2}}$ and $\langle \Delta^0 \rangle = v_\Delta$, with $v^2 + 4 v_\Delta^2 \approx (246 {\rm GeV})^2$, and the minimization conditions are
\begin{align}
    -\mu_H^2 + \lambda_1 v^2 - \frac{1}{2} \mu v_\Delta + \frac{1}{2} \lambda_3 v_\Delta^2 &= 0 \, , \\
    \mu_\Delta^2 + \frac{1}{2} \lambda_3 v^2 - \frac{1}{4}  \frac{\mu}{v_\Delta} v^2 + \frac{1}{2} \lambda_2 v_\Delta^2 &=0 \, ,
    \label{eq:mincond}
\end{align}
which we use to eliminate $\mu_H^2$ and $\mu_\Delta^2$. The scalar mass matrices, in the basis ($H^+, \Delta^+$) and (Re $H^0$, Re $\Delta^0$), are
\begin{align}
   M_+^2 &= \mu \begin{pmatrix}
    v_\Delta & \frac{v}{2} \\
    \frac{v}{2} & \frac{v^2}{4 v_\Delta} 
    \end{pmatrix} \, ,\\
        M_{0}^2 &= 
    \begin{pmatrix}
    2 \lambda_1 v^2 & \frac{-\mu}{2} v + \lambda_3 v\ v_\Delta \\
    \frac{-\mu}{2} v + \lambda_3 v\ v_\Delta & \frac{\mu v^2}{4 v_\Delta} + \lambda_2 v_\Delta^2
    \end{pmatrix}\,,
\end{align}
with mass eigensates
\begin{align}
    G^+ &= \frac{-v H^+ + 2 v_\Delta \Delta^+}{\sqrt{v^2+ 4 v_\Delta^2}}, \hspace{5mm}  \delta^+ = \frac{2 v_\Delta H^+ + v \Delta^+}{\sqrt{v^2+ 4 v_\Delta^2}} \, ,
    \\
  &~~  h = \cos\alpha\ {\rm Re} (H^0) + \sin\alpha\ {\rm Re} (\Delta^0) \, ,\\
   &~~ \delta^0 = -\sin\alpha\ {\rm Re}(H^0) + \cos\alpha\ {\rm Re} (\Delta^0) \, ,
\end{align}
where
\begin{equation}
    \sin 2 \alpha = \frac{\mu v -2 \lambda_3 v v_\Delta}{m_{\delta^0}^2-m_h^2}\,.
\end{equation}
Note that the massless eigenstate is the Goldstone boson ($G^\pm$), eaten up by the $W^\pm$ gauge boson, while the other combination ($\delta^\pm$) is a physical charged Higgs field. The field $h$ is to be identified as the SM-like Higgs boson of mass 125$\,$GeV and in the limit of a small mixing angle $\alpha$, the splitting between the charged and neutral component of the triplet field is $m_{\delta^+}^2-m_{\delta^0}^2 \simeq v_\Delta (\mu - \lambda_2 v_\Delta)$, i.e.,~both components are nearly degenerate. In the limit $v_\Delta \ll v$ we have $v_\Delta = {\mu v^2}/{(4 \mdzero^2)}$.

As stated in the introduction, the triplet model can be minimally extended by two different vector-like leptons (VLLs) in order to allow for couplings to SM leptons:\\

\noindent I) An $SU(2)_L$ doublet with $Y=-1/2$ ($D$) with Yukawa interactions given by
    \begin{align}
    {\cal L_{\rm Y}^{\rm I}} \!\supset & 
    Y_L^{\rm I} \bar{D}_R \ell_L \Delta + Y_R^{\rm I}\bar{D}_L e_R H  + Y^{\rm I} \bar{D}_L \Delta D_R + \text{h.c.} \, .
\end{align}
II) An $SU(2)_L$ triplet with $Y=-1$ ($T$) with Yukawa interactions given by
    \begin{align}
    {\cal L_{\rm Y}^{\rm II}} \supset & 
     Y_R^{\rm II} {\rm Tr}[\bar{T}_L \Delta] e_R  +  Y_L^{\rm II}  H^\dagger \bar{T}_R \ell_L  + Y^{\rm II}  {\rm Tr}[\bar{T}_L \Delta T_R] + \text{h.c.} \, .
\end{align}
where $\ell$ ($e$) is the SM lepton doublet (singlet) and the VLL $T$ is defined as
\begin{equation}
T = \frac{1}{2}
\begin{pmatrix}
    T^- & \sqrt{2} T^0 \\
    \sqrt{2} T^{--} & -T^-
    \end{pmatrix}\,.
\end{equation}

Integrating out the new VLL at the tree-level leads to the following effective interactions
\begin{align}
\mathcal{L}_\text{eff} =
&\frac{|Y_R^{\rm I}|^2}{2 m_D^2} (H^\dagger i \overset{\leftrightarrow}{D_\mu} H)(\bar{e}_R \gamma^\mu e_R) \nonumber\\
- &\frac{3|Y_L^{\rm II}|^2}{16 m_T^2} (H^\dagger i \overset{\leftrightarrow}{D_\mu} H)(\bar{\ell}_L \gamma^\mu \ell_L) \nonumber\\
+ &\frac{|Y_L^{\rm II}|^2}{16 m_T^2} (H^\dagger i \overset{\leftrightarrow}{D^a_\mu} H)(\bar{\ell}_L \sigma^a \gamma^\mu \ell_L) \\
- &\left( \frac{(Y_L^{\text{I}})^* Y_R^\text{I}}{m_D} + \frac{(Y_L^{\rm II})^* Y_R^\text{II}}{2 m_T} \right) \bar{\ell}_L \Delta H e_R\nonumber\,,
\end{align}
which, after EW breaking modify gauge bosons couplings to leptons, affect Higgs decay to lepton pairs $h\to\ell^+\ell^-$ and induce couplings of the triplet scalar $\Delta$ to leptons.

\begin{figure*}[t]
\centering
\includegraphics[width=0.48\textwidth]{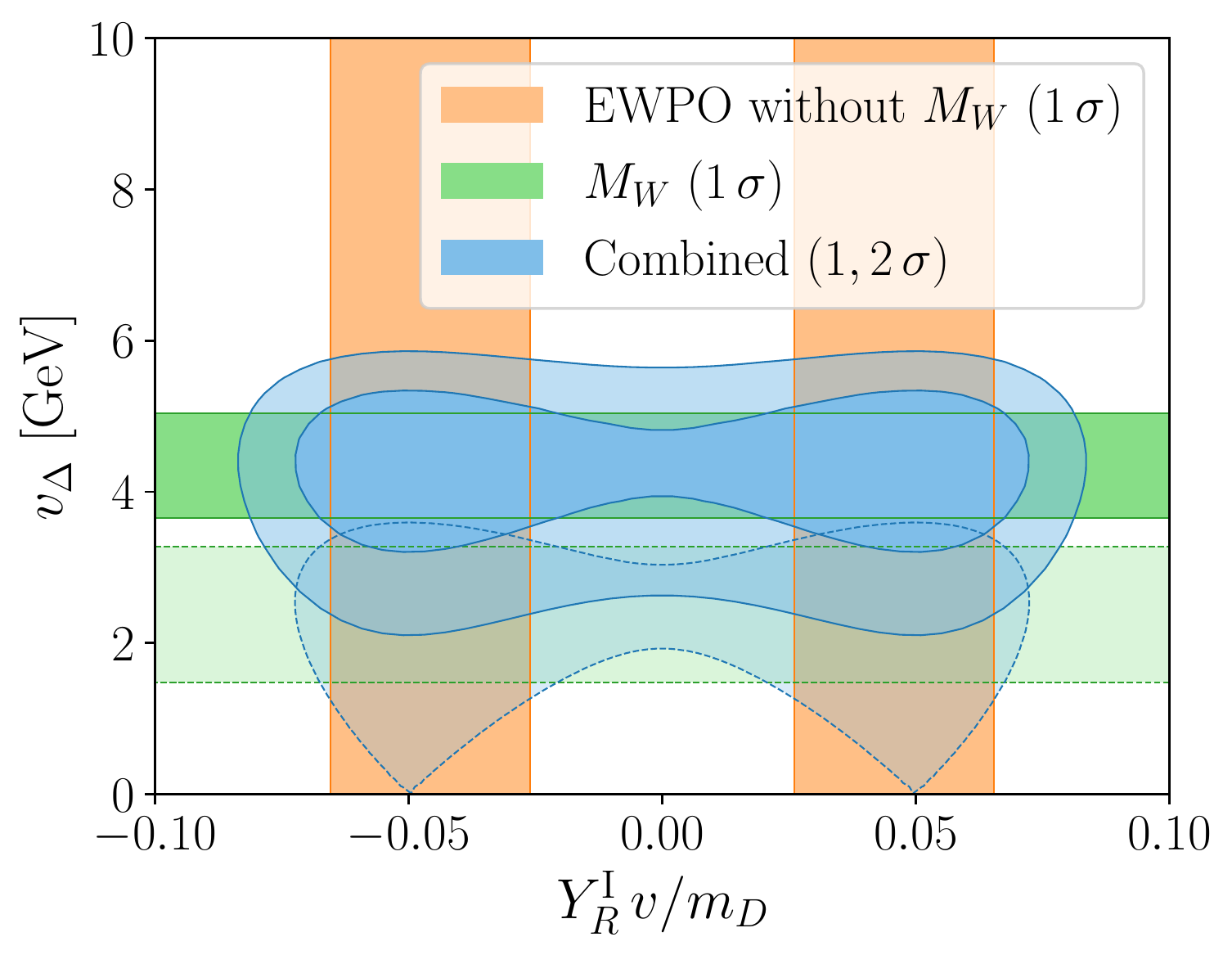}
\hfill
\includegraphics[width=0.48\textwidth]{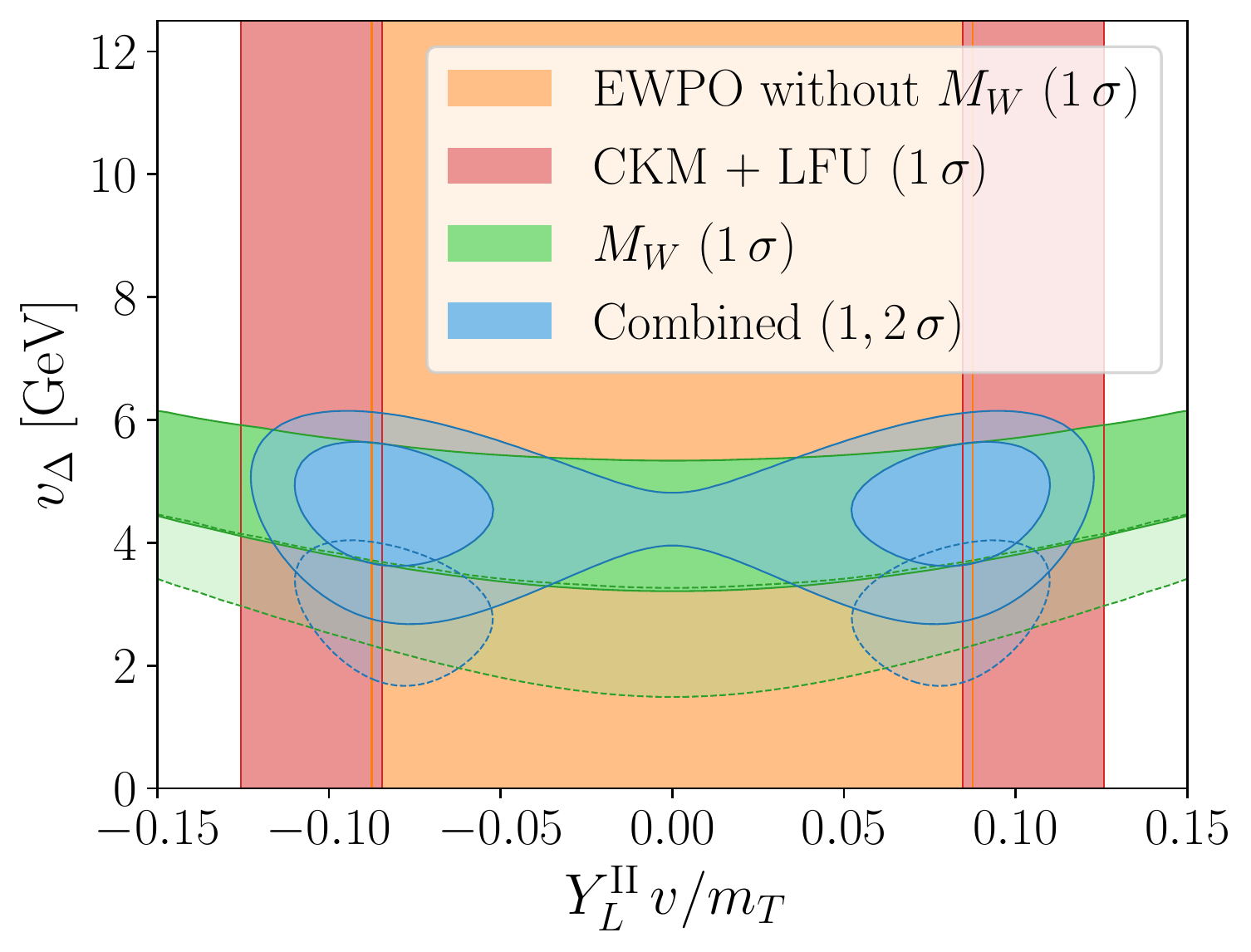}
\caption{Global fit to EW precision data, the $W$ mass, CKM unitarity and tests of LFU, for the case of the VLL doublet (left) and the triplet (right). The preference for a non-zero coupling $Y_R^{\rm I}$ ($Y_L^{\rm II}$) is mainly due to Br$(Z\to\mu^+\mu^-)$ (the CKM unitarity deficit). The solid lines correspond to the conservative average $M_W^{\rm comb}$ while the dashed lines use the $W$ mass average without CDF-II.}
\label{fig:vll_ewpo_mw}
\end{figure*}

\section{Phenomenology}

The CDF~II collaboration updated their previous measurement of the $W$ boson mass, finding $M_W = \SI{80.4335 \pm 0.0094}{\GeV}$, which leads to the new Tevatron average \SI{80.427 \pm 0.0089}{\GeV} when combined with the D0~\cite{CDF:2022hxs} result. However, the recent ATLAS update~\cite{ATLAS:2023fsi} (superseding their 2017 result~\cite{ATLAS:2017rzl}) of $M_W = \SI{80.360 \pm 0.016}{\GeV}$, as well the LHCb~\cite{LHCb:2021bjt}, find significantly smaller values. Together with LEP~\cite{ALEPH:2013dgf}, a naive average gives $M_W = \SI{80.406 \pm 0.007}{\GeV}$. Since the consistency of the data is poor ($\chi^2/\text{dof} = 4.3$), we inflate the error to get a conservative estimate\footnote{Note that these averages agree well with the more sophisticated combinations done by HEPfit~\cite{deBlas:2022hdk} prior to the ATLAS update.} of $M_W^{\rm comb} = \SI{80.406 \pm 0.015}{\GeV}$. Comparing this to the SM prediction of $M_W^\text{SM} = \SI{80.355 \pm 0.005}{\GeV}$~\cite{Sirlin:1983ys,Djouadi:1987gn,Avdeev:1994db,Chetyrkin:1995js,Chetyrkin:1995ix,Awramik:2003rn,Degrassi:2014sxa,ParticleDataGroup:2022pth}, with $m_t = \SI{172.5 \pm 0.7}{\GeV}$~\cite{ParticleDataGroup:2022pth}, we see a discrepancy of \SI{51}{\MeV}, with a significance of slightly more than $\SI{3}{\sigma}$. If instead we disregard the CDF~II result, the data agree well amongst themselves, and we find an average of $M_W^{\rm comb~(w/o~CDF~II)} = \SI{80.372 \pm 0.010}{\GeV}$,
which would correspond to a discrepancy of \SI{17}{\MeV} with a significance of below $2\,\sigma$. In the $\Delta$SM we have
\begin{align}
    m_{W}^2 &= \frac{g^2}{4} (v^2 + 4 v_\Delta^2) \, , \hspace{7mm}  m_Z^2 = \frac{g^2}{4 \cos \theta_W^2} v^2 \, , 
    \label{eq:Wmass}
\end{align}
which shows that the VEV of the triplet can easily alter the $W$ mass in the desired direction\footnote{The triplet scalar $\Delta$ also contributes to the oblique parameters at one loop level, parameterized by the singly charged scalar $m_\delta^+$, the mass difference $\Delta m = m_\delta^+ - m_\delta^0$, and VEV $v_\Delta$. In the limit $v_\Delta << v$, $\Delta m \sim 0$ (required by the perturbative unitarity) and for $m_\delta^+ > 100$ GeV, the one loop contribution is subdominant~\cite{Kanemura:2012rs,Cheng:2022hbo}.
Similarly, the contribution to $S$ and $T$ parameters from the VLLs is suppressed due to VLL mixing with leptons and a loop factor, compared to tree-level effect, which is seen in the very mild to invisible dependence of the $M_W$ regions in Fig.~\ref{fig:vll_ewpo_mw} on the VLL couplings.}. 

Let us start with the tree-level effects induced by the couplings $Y_L^{\rm II}$ and $Y_R^{\rm I}$, which give rise to modifications of EW gauge bosons couplings to leptons. Here, both $D$ and $T$ modify $Z\mu\mu$ couplings, which are constrained from LEP measurements~\cite{ALEPH:2006bhb}, while $T$, in addition, modifies the leptonic $W$ vertex as $\bar{\mu} \gamma^\alpha P_L \nu_\mu W^-_\alpha \to [1 + v^2 |Y_L^\text{II}|^2 / (16 m_T^2)] \bar{\mu} \gamma^\alpha P_L \nu_\mu W^-_\alpha$. Therefore, in case II), the extraction of CKM elements is affected by the determination of the Fermi constant $G_F$ from the muon lifetime~\cite{Crivellin:2021njn} (dominantly $V_{ud}$ from beta decays~\cite{Crivellin:2021njn}).
Furthermore, lepton flavour universality (LFU) measurements in the charged current (see Ref.~\cite{Bryman:2021teu} for an overview) receive new physics contributions. In fact, the CAA, i.e.~the deficit in the first row unitarity relation $|V_{ud}|^2 + |V_{us}|^2 +|V_{ub}|^2 = 1$~\cite{Belfatto:2019swo,Grossman:2019bzp,Crivellin:2020ebi,Coutinho:2019aiy,Crivellin:2020lzu}, with a significance at the $3\,\sigma$ level~\cite{Hardy:2020qwl,Seng:2021nar,Seng:2020wjq,Cirigliano:2022yyo}, can be resolved by the VLL triplet~\cite{Endo:2020tkb,Crivellin:2020ebi,Kirk:2020wdk}. Performing a combined fit using \texttt{smelli v2.4.0}~\cite{Aebischer:2018iyb,smelli_2_4_0}\footnote{For the complete list of observables included in our global fit, we refer the interested reader to Ref.~\cite{Crivellin:2022rhw}, to which we added the tests of LFU $\mathrm{Br}(\tau \to e \nu \nu)$, $\mathrm{Br}(\tau \to \mu \nu \nu)$, $\mathrm{Br}(\pi^+ \to e \nu)$, and $\mathrm{Br}(K^+ \to e \nu) / \mathrm{Br}(K^+ \to \mu \nu)$.} (which is built on \texttt{flavio v2.5.4}~\cite{Straub:2018kue,flavio_2_5_4} for the observable calculations and \texttt{wilson}~\cite{Aebischer:2018bkb} for the renormalization group evolution), we show the region in parameter space favoured by the global EW fit, tests of LFU, the $W$ mass and CKM unitarity in Fig.~\ref{fig:vll_ewpo_mw}. The best-fit points are
\begin{align}
{Y_R^{\rm I} v}/{m_D} &= \pm 0.05, \quad v_\Delta = 4.5\,\text{GeV}\,, \label{eq:best_fit_I} \\
{Y_L^{\rm II} v}/{m_T} &= \pm 0.09, \quad v_\Delta = 4.8\, \text{GeV}\,.
\end{align}
with pulls relative to the SM of \SI{3.1}{\sigma} and \SI{3.6}{\sigma}, respectively (taking into account two degrees of freedom). Note that the preference for a non-zero coupling $Y_R^{\rm I}$ ($Y_L^{\rm II}$) is mainly due to Br$(Z\to\mu^+\mu^-)$ (the CKM unitarity deficit).

For $g-2$ of the muon, the experimental value~\cite{Bennett:2006fi,Abi:2021gix} deviates from the SM prediction ~\cite{Aoyama:2012wk,Aoyama:2019ryr,Czarnecki:2002nt,Gnendiger:2013pva,Davier:2017zfy,Keshavarzi:2018mgv,Colangelo:2018mtw,Hoferichter:2019mqg,Davier:2019can,Keshavarzi:2019abf,Kurz:2014wya,Melnikov:2003xd,Masjuan:2017tvw,Colangelo:2017fiz,Hoferichter:2018kwz,Gerardin:2019vio,Bijnens:2019ghy,Colangelo:2019uex,Blum:2019ugy,Colangelo:2014qya}, resulting in a  $4.2\sigma$ tension
\begin{equation}
\label{Deltaamu}
 \Delta a_\mu[e^+e^-]^{\rm WP}=a_\mu^{\rm exp}-a_\mu^{\rm SM}[e^+e^-]=251(59)\times 10^{-11}, 
\end{equation}
according to the White Paper~\cite{Aoyama:2020ynm}. However, the significance crucially depends on the value used for hadronic vacuum polarization (HPV). While $e^+e^-$ data underlies Eq.~\eqref{Deltaamu}, this dispersive approach has been challenged by lattice QCD~\cite{Borsanyi:2020mff,RBC:2018dos,Ce:2022kxy,ExtendedTwistedMass:2022jpw,Bazavov:2023has,Blum:2023qou,Colangelo:2022vok}, leading to a smaller tension with experiment. The reason for this mismatch is not understood, and also the recent measurement of $e^+e^-\to\pi^+\pi^-$ by CMD-3~\cite{CMD-3:2023alj} differs from previous measurements~\cite{CMD-2:2006gxt,Achasov:2006vp,BaBar:2012bdw,KLOE-2:2017fda,BESIII:2015equ,SND:2020nwa} at a combined level of $5\sigma$. 
Therefore, in this article, we consider ourselves agnostic to the exact value and do not aim for any specific range, merely noting two possible options in the figure to guide the reader.

Neglecting scalar mixing, which is naturally small given the preferred range of $v_\Delta$, the leading (chirally enhanced) 1-loop contribution to the anomalous magnetic moment is given by
\begin{align}
\Delta a_\mu^{\rm I}  = &\frac{m_\mu v (Y_L^{\rm I})^* Y_R^{\rm I} }{64 \sqrt{2} \pi^2 m_D^2 (r-1)^3} \notag
\\
& \times \Big( 
\frac{4 r v_\Delta m_D}{v^2}  \left[ 7 + r (r-8) + (r+2) \log r^2 \right]  \notag \\
& +Y^{\rm I}  \left[ 1+ r (4-5r) + r (r+2 ) \log r^2 \right] \Big)\,,
\\
\Delta a_\mu^{\rm II}  = &\frac{m_\mu v (Y_L^{\rm II})^*  Y_R^{\rm II} }{128 \sqrt{2} \pi^2 m_T^2 (r-1)^3} \notag
\\
\times &\Big( 
\frac{4 r v_\Delta m_T}{v^2} \left[ -1 + r (8-7r) + (5r -2) \log r^2 \right] \notag\\
&- 2 Y^{\rm II}  (r-1) \left[ 1-r + r \log r \right] \Big)\,, 
\end{align}
for the two cases,\footnote{We confirmed these results using \texttt{MatchMakerEFT}~\cite{Carmona:2021xtq}.} where $r = \mdzero^2 / m_{D,T}^2$ for the doublet and triplet case, respectively. The dominant modification of $h \to \mu^+ \mu^-$ arises already at 
tree-level resulting in
\begin{align}
\frac{\text{Br}(h \to \mu \mu)}{\text{Br}(h \to \mu \mu)^\text{SM}} = \left| 1 + \frac{v v_\Delta Y_L Y_R}{N \sqrt{2} m_\mu m_\psi} \right|^2\,,
\end{align}
where $m_\psi = m_D$ or $m_T$ and $N = 1$ or $2$ for the doublet or triplet VLL cases, respectively. The average of the ATLAS~ \cite{ATLAS:2020fzp} and CMS~\cite{CMS:2020xwi} measurements is
\begin{equation}
\frac{\text{Br}(h \to \mu \mu)}{\text{Br}(h \to \mu \mu)^\text{SM}} = 1.21^{+0.36}_{-0.34}\,,
\end{equation}
while a precision of around \SI{10}{\percent} is expected at the HL-LHC with an integrated luminosity of~\SI{3000}{\per\femto\barn}~\cite{Cepeda:2019klc}.

\begin{figure*}[t]
    \centering
    \includegraphics[width=0.48\textwidth]{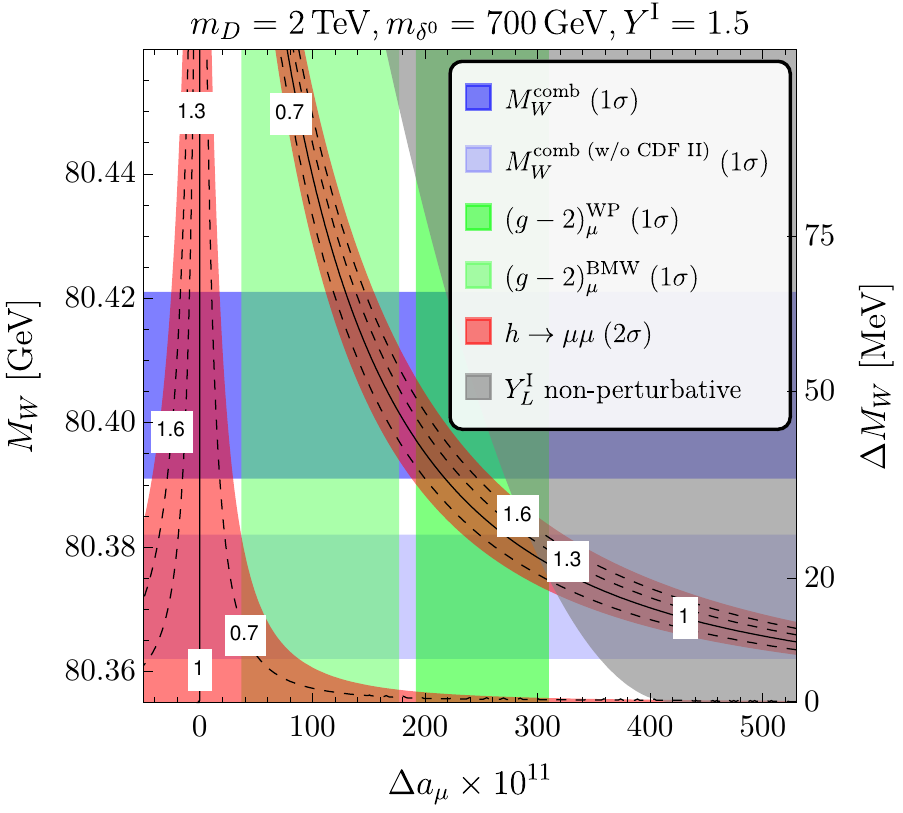} 
\hfill
\includegraphics[width=0.48\textwidth]{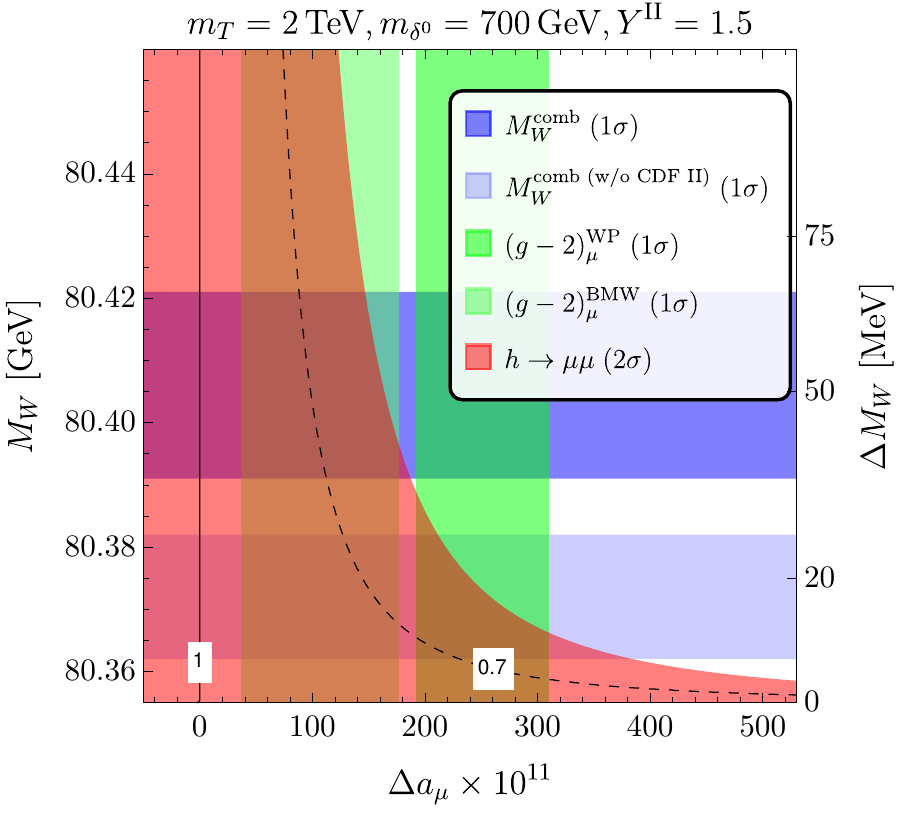}
    \caption{Prediction for $h \to \mu^+ \mu^-$ as a function of $\Delta a_\mu$ and $\Delta M_W$ for model I) (left) and model II) (right). In the left figure, ``$Y_L^\text{I}$ perturbativity'' means $Y_L^\text{I} > 2$, see main text for further details.}
    \label{fig:gm2_vs_mw}
\end{figure*}

Concerning direct LHC bounds, the lower limits on the masses of VLLs which are triplets or doublets of $SU(2)_L$ are around 700$\,$GeV for third generation VLLs~\cite{CMS:2019hsm}, meaning that we expect somehow stronger limits for second generation VLLs and to be conservative we will set the mass to $2\,$TeV. Furthermore, since the VLLs induce couplings of the triplet to muons and muon neutrinos, to a good approximation, the bounds on slepton searches in the limit of a vanishing neutralino mass apply for the mass of the scalar triplet, which are as well around 700$\,$GeV~\cite{CMS:2023qhl}. Taking into account these constraints, in Fig.~\ref{fig:gm2_vs_mw} we predict Br$(h \to \mu^+ \mu^-)$ as a function of $(g-2)_\mu$ and $M_W$. 
Note that $\Delta a_\mu$ can be as large as $250 \times 10^{-11}$ while at the same time providing a sizable effect in $M_W$.
It is important to note that the numerical value of the loop function entering $(g-2)_\mu$ is larger in model II) than in model I). Therefore, in model II) one can obtain a sizable effect with smaller couplings $Y_{L,R}$ which leads to the small effects in $h \to \mu \mu$ compared to model I), as well as to bounds from the perturbativity of $Y_L^\text{I}$ (if $Y_R^\text{I}$ is fixed to the best-fit value in Eq.~(\ref{eq:best_fit_I}).

\section{Conclusions}

In this article, we proposed (two versions of) a minimal model obtained by extending the SM with a scalar triplet with hypercharge 0 and a vector-like lepton that is I) an $SU(2)_L$ doublet with $Y=-1/2$ or II) $SU(2)_L$ triplet with $Y=-1$. This model can:
\begin{itemize}
    \item Provide naturally a positive definite shift in the $W$ mass of the size suggested by the current tension.
    \item Give a sizable effect in $g-2$ of the muon.
    \item Improve on $Z\to\mu^+\mu^-$ in case I) or explain the CAA in case II).
\end{itemize}
For both model versions, effects in $h\to\mu^+\mu^-$ of the order of 10\% (while most other models on the market only generate $O(\%)$ effects), as well as $\mu^+\mu^-$ plus missing energy signatures at the LHC, are predicted.

\begin{acknowledgments}
A.C.~thanks Martin Hoferichter for useful discussions concerning the status of the SM prediction for $(g-2)_\mu$. Financial support from the SNSF (PP00P21\_76884) is gratefully acknowledged. M.K.~acknowledges support from a Maria Zambrano fellowship, and from the State Agency for Research of the Spanish Ministry of Science and Innovation through the ``Unit of Excellence Mar\'ia de Maeztu 2020-2023'' award to the Institute of Cosmos Sciences (CEX2019-000918-M) and from PID2019-105614GB-C21 and 2017-SGR-929 grants. The work of A.T.~is supported in part by the National Science Foundation under Grant PHY-2210428. A.T.~acknowledges the Department of Physics at Washington University in St.~Louis for local hospitality during the completion of this work. 
\end{acknowledgments}

\bibliographystyle{utcaps_mod}
\bibliography{references}
\end{document}